\documentclass{emulateapj}

\usepackage{amsmath,amssymb,latexsym,graphics}

\shorttitle{\textsc{Time Delay and Microlensing in SBS\,0909+532}}
\shortauthors{\textsc{Hainline et al.}}

\newcommand{\kms}{\textrm{km\,s}^{-1}}
\newcommand{\mpc}{\textrm{Mpc}}
\newcommand{\msun}{M_{\sun}}

\begin{document}

\title{Time Delay and Accretion Disk Size Measurements in the Lensed Quasar SBS\,0909+532 
from Multiwavelength Microlensing Analysis}

\author{Laura J.\ Hainline\altaffilmark{1}, Christopher W.\ Morgan\altaffilmark{1}, 
Chelsea L.\ MacLeod\altaffilmark{1}, Zachary D.\ Landaal\altaffilmark{1},
C. S.\ Kochanek\altaffilmark{2}, Hugh C.\ Harris\altaffilmark{3}, Trudy Tilleman\altaffilmark{3},
L.\ J.\ Goicoechea\altaffilmark{4},  V.\ N.\ Shalyapin\altaffilmark{4,5}, and Emilio E.\ Falco\altaffilmark{6} }

\altaffiltext{1}{Department of Physics, United States Naval Academy,
572C Holloway Rd, Annapolis, MD 21402, USA; hainline@usna.edu,
cmorgan@usna.edu, macleod@usna.edu, m123894@usna.edu}
\altaffiltext{2}{Department of Astronomy, The Ohio State University,
140 West 18th Ave, Columbus, OH 43210, USA; ckochanek@astronomy.ohio-state.edu}
\altaffiltext{3}{United States Naval Observatory, Flagstaff Station,
10391 West Naval Observatory Road, Flagstaff, AZ 86001-8521, USA;
hch@nofs.navy.mil, trudy@nofs.navy.mil}
\altaffiltext{4}{Facultad de Ciencias, Universidad de Cantabria, Avda. de Los Castros s/n, 
39005 Santander, Spain; goicol@unican.es}
\altaffiltext{5}{Institute for Radiophysics and Electronics, National Academy of Sciences 
of Ukraine, 12 Proskura St., 61085 Kharkov, Ukraine; vshal@ukr.net}
\altaffiltext{6}{Harvard-Smithsonian Center for Astrophysics, 60 Garden St, Cambridge, MA 02138, USA;
falco@cfa.harvard.edu}

\begin{abstract}

We present three complete seasons and two half-seasons of SDSS $r$-band photometry of the 
gravitationally lensed quasar SBS\,0909+532 from the U.S. Naval Observatory, as well as 
two seasons each of SDSS $g$-band and $r$-band monitoring from the Liverpool Robotic Telescope.  
Using Monte Carlo simulations to simultaneously measure the system's time delay
and model the $r$-band microlensing variability, we confirm and significantly refine
the precision of the system's time delay to $\Delta t_{AB} = 50^{+2}_{-4}\,\textrm{days}$, where 
the stated uncertainties represent the bounds of the formal $1\,\sigma$ confidence interval. 
There may be a conflict between the time delay measurement and a
lens consisting of a single galaxy.  While models based on the
{\it Hubble Space Telescope} astrometry and a relatively compact stellar distribution can
reproduce the observed delay, the models have somewhat less dark matter
than we would typically expect.  We also
carry out a joint analysis of the microlensing variability in the $r$- and $g$-bands to constrain the
size of the quasar's continuum source at these wavelengths, obtaining 
$\log \{(r_{s,r}/\textrm{cm})[\cos{i}/0.5]^{1/2}\} = 15.3 \pm 0.3$ and 
$\log \{(r_{s,g}/\textrm{cm})[\cos{i}/0.5]^{1/2}\} = 14.8 \pm 0.9$, respectively.  
Our current results do not formally constrain the temperature profile of the accretion disk 
but are consistent with the expectations of standard thin disk theory.
\end{abstract}


\section{INTRODUCTION}\label{sec:intro}

Much of the standard picture for the detailed structure of accretion disks surrounding
supermassive black holes in active galactic nuclei is based on 
theoretical models rather than observational measurements, because these very compact regions 
cannot be resolved with existing telescopes.
Yet, for gravitationally lensed quasars, the relative motions of the observer,
the background source, the foreground lens galaxy, and its stars cause
uncorrelated variations in the source magnification as a function of time and wavelength which depend on 
the projected area of the continuum source.  By analyzing these microlensing brightness fluctuations
with numerical simulations, one can measure the continuum source size, 
permitting invaluable observational tests of theoretical models of accretion 
disk structure \citep[e.g.,][]{eigenbrod08,anguita08,poindexter08} and orientation
\citep{poindexter10}.  Such tests in the literature typically fall into two categories.
One type of study employs single-epoch multi-band photometry of
lensed quasars in which flux ratios of the images exhibit deviations from the 
predictions of macroscopic lens models or significant wavelength dependence
\citep[e.g.,][]{pooley06,bate08,blackburne11,mosquera11,motta12}.  The alternative method, 
described in detail in \citet{kochanek04}, analyzes the time variability 
of the quasar's flux ratio and requires monitoring a quasar over a significant period of time.  Although
such investigations can be observationally and computationally challenging, they have the advantage that
it is not necessary to assume a value for the mass of the microlenses or an extinction
law for the lens galaxy, nor are the results highly sensitive to assumed priors.
Studies employing both methods of analysis have revealed accretion disk temperature profiles 
in agreement with the simple thin-disk model of \citet{shakura73}, although
the individual observed disk sizes tend to be larger than those predicted by the theoretical model
\citep[e.g.][]{pooley07,morgan10,blackburne11}.

SBS\,0909+532 (hereafter SBS\,0909; $\alpha_{\textrm{J}2000} = 09^{\textrm{h}}13^{\textrm{m}}01\fs05$, 
$\delta_{\textrm{J}2000} = +52^{\textrm{d}}59^{\textrm{m}}28\fs83$) 
is a doubly-imaged quasar lens system in which the background
quasar has redshift $z_{s} = 1.377$ and the foreground early-type lens galaxy
has redshift $z_{l} = 0.830$ \citep{kochanek97,oscoz97,lubin00}.  SBS\,0909
is a somewhat challenging system to study because discrepant results in the literature 
expose significant uncertainties about some of its fundamental properties and 
the nature of the quasar's variability. Most notably, 
\citet{lehar00} encountered difficulty when
attempting to use the {\it imfitfits} routine to measure the lens galaxy photometry in
$H$-band NICMOS images from the \emph{Hubble Space Telescope} (\emph{HST}),
eventually settling on a low surface brightness de Vaucouleurs model with effective 
radius $r_ {\textrm{eff}} = 1\farcs58\pm0\farcs9$ and magnitude
$H = 16.75$.  In stark contrast, \citet{sluse12} used an iterative deconvolution technique
\citep[e.g.,][]{magain98,chantry07} on the same data to find a significantly
smaller ($r_{\textrm{eff}}=0\farcs54\pm0\farcs02$) and less luminous ($H=19.44\pm0.01$) lens 
galaxy, formally inconsistent with the \citet{lehar00} result.
Additionally, past optical monitoring observations have not 
shown evidence of significant microlensing variability \citep{ullan06,luis08}, although
\citet{mediavilla11} find evidence of chromatic microlensing (wavelength-dependent microlensing
magnification) through an analysis of the differences between the continuum and emission line flux ratios 
from the quasar's ultraviolet (UV) -- near-infrared (near-IR) spectra, which can separate microlensing
effects from differential extinction present in the lens galaxy.  
\begin{figure}
\begin{center}
\epsscale{1.2}
\plotone{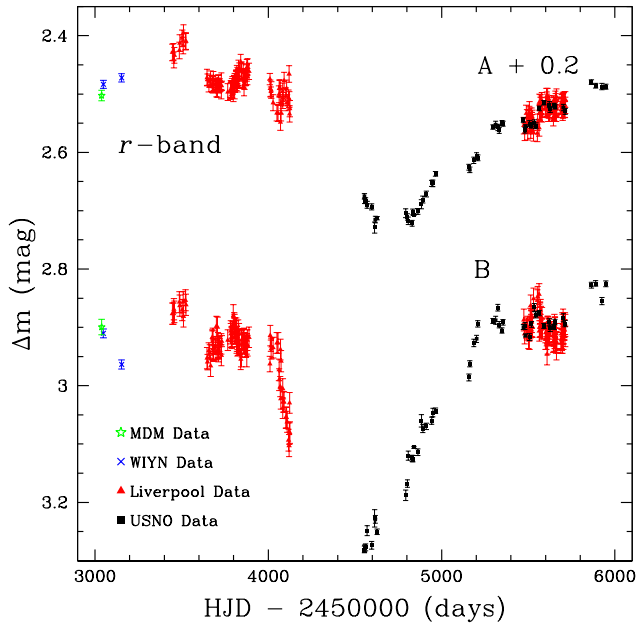}
\caption{Composite $r$-band light curves for SBS\,0909 images A (top) and B (bottom) including 
measurements from MDM Observatory (stars), the WIYN 3.5-m telescope (diagonal crosses), 
the Liverpool Telescope (triangles), and USNO (squares).  The measurements for 
image A have been offset by +0.2\,mag to minimize empty space in the plot area.   The light 
curve of image B exhibits a substantially steeper slope over the time period 
$4500 \lesssim \textrm{HJD} - 2450000 \lesssim 5200\,\textrm{days}$. \label{fig:r_lightcurve}}
\end{center}
\end{figure}
\begin{deluxetable*}{lcccc}
\centering
\tablewidth{0pt}
\tablecolumns{5}
\tablecaption{SBS\,0909+532 Light Curves from USNO\label{tab:lightcurve}}
\tablehead{
\colhead{HJD - 2450000} & \colhead{Seeing} & \colhead{QSO A} & \colhead{QSO B} & \colhead{$\langle \textrm{Stars} \rangle$} \\
\colhead{(days)}         & \colhead{(arcsec)} & \colhead{(mag)} & \colhead{(mag)} & \colhead{(mag)}
}
\startdata
$4554.635$ & $1.2$ & $ 2.475\pm 0.005$ & $ 3.281\pm 0.006$ & $ 0.033\pm 0.003$ \\ 
$4555.635$ & $1.2$ & $ 2.483\pm 0.005$ & $ 3.279\pm 0.006$ & $ 0.034\pm 0.003$ \\ 
$4561.686$ & $1.7$ & $ 2.483\pm 0.005$ & $ 3.277\pm 0.006$ & $ 0.031\pm 0.003$ \\ 
$4570.625$ & $1.5$ & $ 2.490\pm 0.006$ & $ 3.249\pm 0.009$ & $-0.087\pm 0.003$ \\ 
($4584.638$) & ($1.9$) & $( 2.494\pm 0.005)$ & $( 3.245\pm 0.006)$ & $( 0.008\pm 0.003)$ \\ 
$4596.651$ & $1.3$ & $ 2.493\pm 0.005$ & $ 3.273\pm 0.007$ & $-0.016\pm 0.003$ \\ 
$4613.664$ & $1.6$ & $ 2.529\pm 0.009$ & $ 3.227\pm 0.015$ & $-0.061\pm 0.004$ \\ 
$4617.673$ & $1.5$ & $ 2.515\pm 0.006$ & $ 3.229\pm 0.007$ & $ 0.014\pm 0.003$
\enddata
\tablecomments{HJD is the Heliocentric Julian Day.  The magnitudes listed in the QSO A and B columns are measured
relative to the comparison stars.  The magnitudes in the $\langle \textrm{Stars} \rangle$
column are the mean magnitudes of the comparison stars for that epoch relative to
their mean over all epochs.  The light curve points listed in parentheses have not been included
in the analysis. Table 1 will be published in its entirety in the electronic edition of {\it The Astrophysical
Journal}.  A portion is shown here for guidance regarding its form and content.}

\end{deluxetable*}

We have compiled a new data set consisting of monitoring observations 
of SBS\,0909 in two optical bands from two different observatories, the analysis of which
provides some resolution to these discrepant results from the literature.  
Here we will show that in the
four years spanning 2008 -- 2012, SBS\,0909 has exhibited significant uncorrelated time variability in
the rest-frame near-UV.  We will analyze the uncorrelated variability, which we attribute
to microlensing, to determine a size for the accretion disk in the two different bands, 
and then compare our results to that derived from the observed chromatic microlensing by \citet{mediavilla11}.  
Since our multi-band data set allows us to constrain the size of the accretion disk at two different
wavelengths, we also gain a glimpse at the temperature profile of the accretion disk, 
with less reliance on priors than single-epoch microlensing analyses.  

In order to analyze the uncorrelated variability of a lensed quasar using the methods
of \citet{kochanek04}, we must first have accurate knowledge of the time delay between 
the lensed images.  In general, imprecise time delays
can result in significant uncertainties in microlensing analyses, since residual 
variability from an improperly corrected time delay must be modeled as uncorrelated microlensing variability in 
simulated light curves.  In the case of SBS\,0909,  the lens mass model provides very little help with reducing the time delay uncertainty 
given the very discrepant lens galaxy photometric fits of \citet{lehar00} and \citet{sluse12}.
Unfortunately, the time delay of SBS\,0909 has been rather difficult
to determine because there have been relatively few large-amplitude ($\ga 0.1$~mag), short-duration
extrema in the light curves in the years since its discovery. 
The most recent time delay measurement for this quasar, $\Delta t_{AB} = 49 \pm 6$\,days \citep{luis08},
still had relatively significant uncertainty.  
In our new $r$-band data set for SBS\,0909, the quasar 
images exhibit substantial intrinsic flux variation, but this flux variation is
modulated by the uncorrelated microlensing variability so that a simple polynomial-based 
cross-correlation analysis \citep[e.g,][]{kochanek06,Poindexteretal.2007} 
fails, a problem explored by \citet{Eigenbrod2005}. \citet{morgan08} introduced a method of making this problem
tractable, by analyzing the microlensing variability with the techniques from \citet{kochanek04}
while simultaneously solving for the time delay.  We will apply this methodology to our
$r$-band data set to make a new, independent measurement of the time delay for SBS\,0909 which
provides strong evidence in favor of a more compact lens galaxy photometric model \citep[e.g.][]{sluse12} but at the astrometric 
position favored by \citet{lehar00}.
\begin{figure}
\begin{center}
\epsscale{1.2}
\plotone{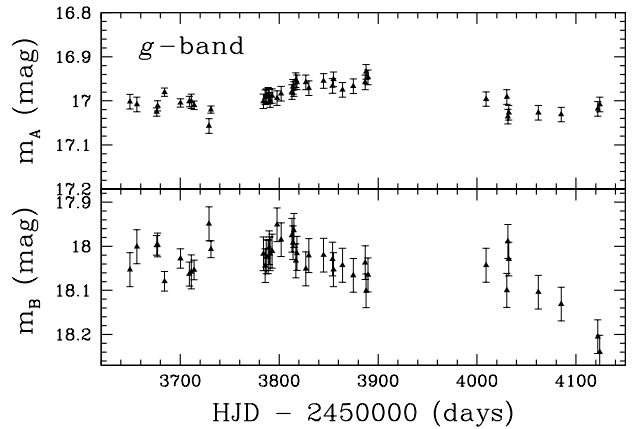}
\caption{Observed $g$-band light curves for SBS\,0909 images A (top panel) and B (bottom
panel) from the Liverpool Telescope.  The $g$-band light curves 
exhibit similar intrinsic variability to the $r$-band light
curves over the same period of time, although with increased scatter due to the lower quasar
flux in $g$ and poorer observing conditions on some occasions. \label{fig:g_lightcurve}}
\end{center}
\end{figure}

In Section~\ref{sec:obs_data} we describe our $g$- 
and $r$-band monitoring observations from the United States Naval Observatory -- Flagstaff
and the Liverpool Telescope and how we combined the two data sets.  In Section~\ref{sec:analysis_method}
we outline our Bayesian Monte Carlo method for simultaneously estimating time delays
and quasar structure.  In Section~\ref{sec:results_disc} we present the results of our analysis
and compare our findings to those of \citet{mediavilla11}.  Throughout our
discussion we assume a flat cosmology with $\Omega_{\textrm{M}}=0.3$,
$\Omega_{\Lambda}=0.7$, and $H_{0}=70\,\kms~\mpc^{-1}$ \citep{hinshaw09}.

\section{OBSERVATIONAL DATA}\label{sec:obs_data}

\subsection{USNO Monitoring}

We observed SBS\,0909 regularly as part of the United States Naval Academy/United States Naval 
Observatory (USNA/USNO) Lensed Quasar Monitoring Program.
Using the 1.55-m Kaj Strand Astrometric Reflector at the USNO -- Flagstaff Station, 
we take three five-minute exposures of the quasar in the Sloan
Digital Sky Survey (SDSS) $r$-band per epoch at a cadence of 
two to three nights per month, using either
the $2048 \times 2048$ Tek2K CCD camera ($0\farcs33\,\textrm{pixel}^{-1}$) or the
$2048 \times 4096$ EEV CCD camera ($0\farcs18\,\textrm{pixel}^{-1}$).
The details of our photometric analysis techniques are 
discussed in detail in \citet{kochanek06}.  In summary, we measure the quasar image 
fluxes relative to three reference stars, located at $(-12\farcs8, 71\farcs9)$, 
$(67\farcs2, 11\farcs4)$, and $(-13\farcs8, 7\farcs1)$ with respect to image A of 
SBS\,0909, using a three-component elliptical Gaussian point-spread function (PSF) model.  
We hold the relative positions of the quasar images fixed to those derived from the 
\emph{HST} $H$-band images of SBS\,0909 for the 
PSF fitting process.  The photometric model
of the very red $z_{l}=0.83$ lens galaxy is a Gaussian approximation to a de Vaucouleurs profile
of fixed effective radius and flux.  We use the effective radius
derived from the \emph{HST} images by \citet{lehar00} and for the flux we use the
value which minimizes the total $\chi^{2}$ in the residuals over all epochs. 
We also attempted to measure the quasar photometry using 
the more compact and dimmer lens galaxy photometric fit from \citet{sluse12},
but the changes in the quasar image fluxes were negligible.
Unlike the case of Q\,0957+561 in \citet{hainline12}, no color offset is required
between the two different detectors used for our observing program.   
In Table~\ref{tab:lightcurve}, we present the $r$-band measurements of SBS\,0909 components
A and B from 61 nights between 2008 March and 2012 February.  The
images from which our measurements are derived are characterized by a 
median stellar FWHM (seeing) of $1\farcs 3$.
Because the quasar images are closely spaced ($1\farcs17$),
they are blended in our USNO images, causing our photometric analysis to break down for
seeing conditions somewhat larger than the image separation, so we keep only 
epochs for which the seeing is better than $1\farcs6$ in our analysis.  
This removes 10 epochs from our 
USNO data set, which are identified in Table~\ref{tab:lightcurve} by the parentheses
surrounding the measurements.  We discarded an additional 6 epochs
not listed in Table~\ref{tab:lightcurve} due to partial cloud cover and bright sky
conditions.
\begin{deluxetable}{lcc}
\tabletypesize{\scriptsize}
\centering
\tablewidth{0pt}
\tablecolumns{3}
\tablecaption{SBS\,0909+532 $r$ Light Curves from the Liverpool Telescope \label{tab:lt_r_lightcurve}}
\tablehead{
\colhead{HJD - 2450000} & \colhead{QSO A} & \colhead{QSO B}  \\
\colhead{(days)}         & \colhead{(mag)} & \colhead{(mag)} 
}
\startdata
4009.709  & $16.379 \pm 0.010$  & $17.073 \pm 0.013$ \\
4011.697  & $16.394 \pm 0.014$  & $17.019 \pm 0.018$ \\
4015.729  & $16.394 \pm 0.008$  & $17.053 \pm 0.010$ \\
4017.705  & $16.397 \pm 0.014$  & $17.042 \pm 0.018$ \\
4028.658  & $16.419 \pm 0.014$  & $17.040 \pm 0.018$ \\
4029.668  & $16.416 \pm 0.014$  & $17.156 \pm 0.018$ \\
4030.708  & $16.403 \pm 0.014$  & $17.117 \pm 0.018$ \\
4031.654  & $16.393 \pm 0.014$  & $17.097 \pm 0.018$ 
\enddata
\tablecomments{HJD is the Heliocentric Julian Day.  The magnitudes listed
have been calibrated to the SDSS photometric system using the flux of the ``b'' star 
in the field of SBS\,0909 \citep[see][]{kochanek97}. Table 2 will be published in its entirety in the electronic edition of {\it The Astrophysical
Journal}.  A portion is shown here for guidance regarding its form and content.}

\end{deluxetable}

\begin{deluxetable}{lcc}
\tabletypesize{\scriptsize}
\centering
\tablewidth{0pt}
\tablecolumns{3}
\tablecaption{SBS\,0909+532 $g$ Light Curves from the Liverpool Telescope \label{tab:lt_g_lightcurve}}
\tablehead{
\colhead{HJD - 2450000} & \colhead{QSO A} & \colhead{QSO B}  \\
\colhead{(days)}         & \colhead{(mag)} & \colhead{(mag)}
}
\startdata
3649.708  &  $17.002 \pm  0.017$  &  $18.053 \pm 0.038$ \\
3656.720  &  $17.008 \pm  0.017$  &  $18.001 \pm 0.038$ \\
3676.666  &  $17.025 \pm  0.010$  &  $17.998 \pm 0.022$ \\
3677.665  &  $17.011 \pm  0.012$  &  $17.997 \pm 0.027$ \\
3684.674  &  $16.980 \pm  0.010$  &  $18.079 \pm 0.022$ \\
3700.678  &  $17.005 \pm  0.010$  &  $18.028 \pm 0.022$ \\
3709.702  &  $17.001 \pm  0.012$  &  $18.063 \pm 0.027$ \\
3711.699  &  $17.001 \pm  0.017$  &  $18.058 \pm 0.038$ \\
3714.595  &  $17.010 \pm  0.010$  &  $18.054 \pm 0.022$
\enddata
\tablecomments{HJD is the Heliocentric Julian Day.  The magnitudes listed
have been calibrated to the SDSS photometric system using the flux of the ``b'' star
in the field of SBS\,0909 \citep[see][]{kochanek97}. Table 3 will be published in its entirety in the electronic edition of {\it The Astrophysical
Journal}.  A portion is shown here for guidance regarding its form and content.}

\end{deluxetable}

\subsection{Liverpool Telescope Monitoring}

We also monitored SBS\,0909 in the $r$-band with the 2.0-m Liverpool Robotic 
Telescope independently of the observations at USNO.  Our $r$-band monitoring program 
used the RATCam CCD camera, providing a $4\farcm6$ field of view 
with pixel scale $0\farcs28\,\textrm{pixel}^{-1}$, and was carried out over
two different periods: from 2005 January to 2007 January (I), and from 2010 
October to 2012 March (II).  The measurements spanning the time period 2005 January -- 2006 June 
(78 epochs) have already been published in \citet{luis08}.  Here we add 30 additional epochs
of magnitudes corresponding to the last time 
segment from monitoring period I (2006 October to 2007 January)
and the first 90 epochs from monitoring period II (2010 October to 2011 June), for a total
of 198 epochs on the Liverpool Telescope. 

In addition, we present here data from the $g$-band monitoring program of SBS\,0909 at the 
Liverpool Telescope, contemporaneous with the $r$-band data set published
in \citet{luis08} and thus spanning two years (2005 January to 2007 January). 
We used the RATCam instrument at the Liverpool Telescope and obtained 167 
individual exposures (frames) of 100 or 200\,s each. After bias subtraction, overscan region trimming, 
and flat fielding of the images, 
a crowded-field PSF photometry pipeline measures instrumental fluxes for bright 
field stars and the quasar images. We then transform the photometry to the SDSS
photometric system, correcting the instrumental fluxes for the systematic effects of 
color and inhomogeneity \citep[see][]{luis10}. The transformation pipeline is 
only applied to the frames in which the signal-to-noise ratio (SNR) of 
the ``c'' field star, measured through an aperture of radius equal to twice the 
FWHM, is greater than 100, and for which the seeing is less than $2\arcsec$.
We also discard frames requiring anomalous color coefficients, frames which
produce photometry outliers, and frames in which the quality of the  
PSF fits to the quasar region are poor. After averaging together 
the photometry from individual frames obtained on the same night,
the final Liverpool Telescope $g$-band data set consists of 43 epochs of SDSS 
magnitudes with average uncertainties of 0.016\,mag (image A) and 0.036\,mag (image B).

The new $r$-band photometry from the Liverpool Telescope is listed in 
Table~\ref{tab:lt_r_lightcurve} and the complete $g$-band data set is
provided in Table~\ref{tab:lt_g_lightcurve}.  We list the Liverpool Telescope light 
curves separately from the USNO light curves because the two data sets
have different photometric calibrations: the Liverpool data frames have been 
calibrated onto an absolute system using the absolute flux of a reference
star, while the USNO measurements are not calibrated to a standard system.  By presenting
the two data sets separately, we preserve the original photometric system
of each and provide transparency of origin for future users of the data sets.

\subsection{Construction of Light Curves and Difference Light Curves}\label{sec:construct_lightcurves}

In order to construct light curves spanning the longest possible time baseline,
we combined all the $r$-band USNO and Liverpool data for SBS\,0909 along with 
one $R$-band epoch obtained at the MDM Observatory's 
Hiltner 2.4-m telescope, using the $1024 \times 1024$ ``Templeton'' CCD camera,
and two SDSS $r$-band epochs obtained with the WIYN Tip-Tilt Module (WTTM) 
at the Wisconsin--Indiana--Yale--NOAO (WIYN) 
3.5-m telescope.  Since these epochs were not contemporaneous with each other or 
our USNO/Liverpool light curves, we were unable to make an empirical measurement
of any magnitude offsets arising from differences between the detectors and filters.  
We accounted for this unknown offset by applying an additional $0.02\,\textrm{mag}$ of
uncertainty to the time-delay corrected flux ratio for these observations.
We determined the magnitude offset between our USNO and Liverpool 
data sets ($-14.108 \pm 0.018$) by making a weighted average of the 
offset $\Delta (\textrm{USNO} - \textrm{LRT})$
found for the 7 individual nights with contemporaneous observations.  

In Figure~\ref{fig:r_lightcurve} we show the combined $r$-band light curves for images 
A and B from all of our data sources.  The dominant feature in the light curves
is the intrinsic variability, as much as 0.3 -- 0.4\,mag, the analysis of which we
will present in a future paper.  Closer inspection, though, of the time range of
$4500 \lesssim \textrm{HJD} - 2450000 \lesssim 5600$ reveals uncorrelated variability,
as the slope of the increase in the brightness of image B over this time period is
noticeably steeper than the slope of image A's light curve.  We attribute this component
of the variability to microlensing by stars in the lens galaxy.  A careful examination of the lightcurves
and Tables~\ref{tab:lightcurve}~and~\ref{tab:lt_r_lightcurve} will reveal that our data cadence is
somewhat variable. In the first few seasons there are many periods with several observations per week, but much of the USNO
data is at a cadence of 1-2 observations per month.  The strength of our analysis technique is that while the sparsely sampled
periods do not constrain the time delay directly, their long time baseline provides very strong constraints
on the microlensing model, thereby severely limiting the number of trial light curves from our Monte Carlo simulation with 
adequate fits to the data during the densely sampled intervals (see Section~\ref{sec:delay}).

We display the shorter observed $g$-band light curves for SBS\,0909, which are 
composed entirely of Liverpool Telescope data, in Figure~\ref{fig:g_lightcurve}.   The intrinsic
variability in the $g$-band is similar to that in the $r$-band during the same time period.  The 
$g$-band curves exhibit considerably more scatter than the $r$-band curves, 
particularly for image B.  The increased scatter is due primarily
to the lower quasar flux in $g$-band and the increased difficulty in extracting
bluer $g$-band fluxes under poor observing conditions.  
Although the increased scatter impedes visual identification of microlensing variability
in the $g$-band light curves, the measurements are still valuable as a complementary 
data set for a simultaneous two-band microlensing analysis. 

\begin{figure}
\begin{center}
\includegraphics[scale=0.43]{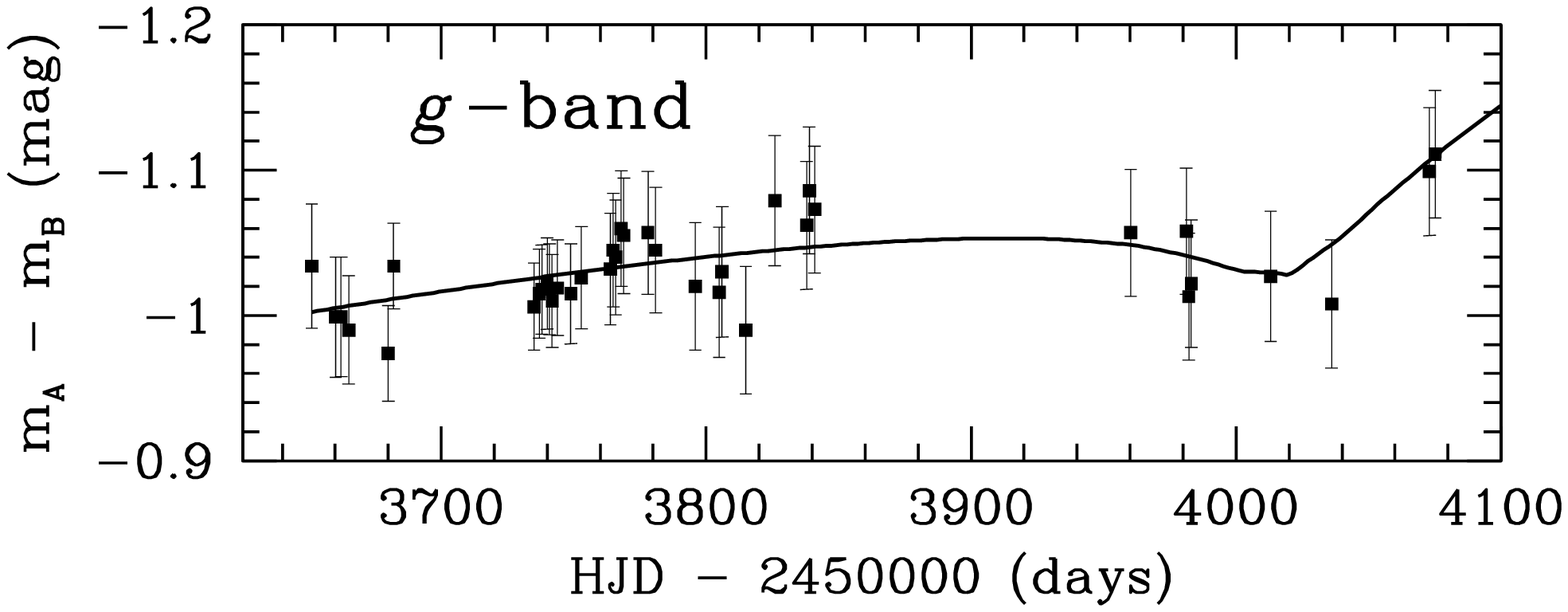}
\includegraphics[scale=0.43]{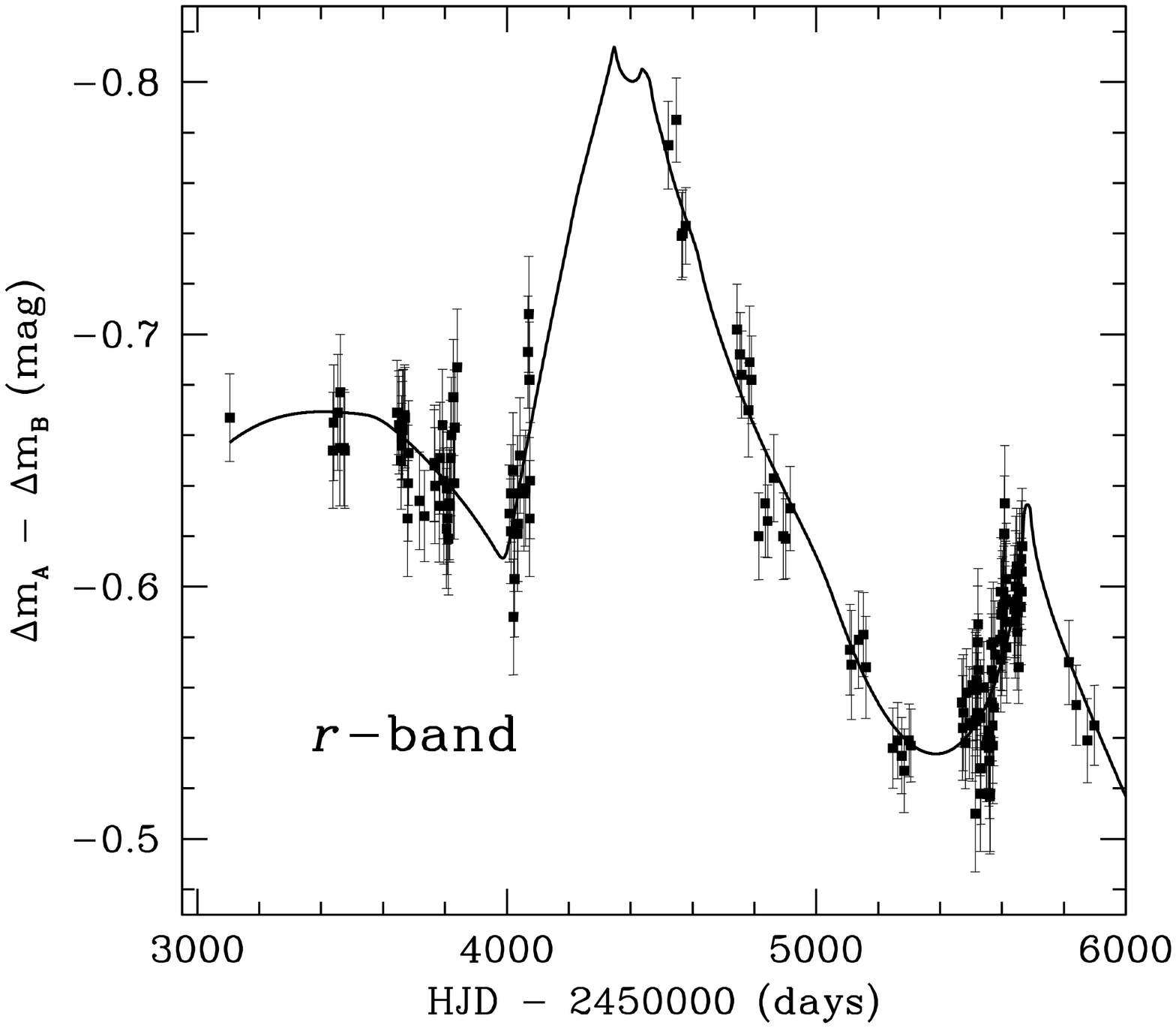}
\caption{Difference light curves in $g$ (top panel) and $r$-band (bottom panel) for SBS\,0909, shown with
an example of a simulated light curve from our Monte Carlo simulations that is a 
good fit to the observations.  To construct the light curves, image A's 
data has been shifted by $\Delta t_{AB} = t_{A} - t_{B}= 50\,\textrm{days}$.
Significant uncorrelated variability is apparent in the $r$-band.\label{fig:diff_lightcurve_fits}}
\end{center}
\end{figure}

The microlensing variability in the $g$- and $r$-band light curves is displayed in 
Figure~\ref{fig:diff_lightcurve_fits}.  These time-delay-shifted difference light curves 
are generated by shifting the light curve of the less variable image 
(image A, in both bands) by the system's 50-day time delay $\Delta t_{AB}$
and then performing a linear interpolation of image A's shifted light
curve to generate a set of photometric measurements at the same epochs of observation
as those in image B's (unshifted) light curve. We discard any data points that
were interpolated in the inter-season gaps.  Finally, we subtract the 
light curve of image B from the shifted and interpolated light curve of image A.  We do not 
apply any corrections for lens galaxy reddening to the light curves or difference light curve.  Rather,
we allow for a 0.5\,mag systematic uncertainty in the intrinsic value flux ratios in our simulations, which 
accounts for the uncertainty in the intrinsic flux ratio from both microlensing and dust extinction 
in the lens galaxy \citep{mediavilla11}.  The resulting difference light curves
are shown in Figure~\ref{fig:diff_lightcurve_fits} for both $g$ and $r$.   As expected,
the microlensing signal is difficult to identify in the shorter and noisier $g$-band difference light curve.
However, the uncorrelated variability in the $r$-band is quite clear.  We observe a steep
fall and rise in $\Delta m_{A} - \Delta m_{B}$ in the time interval 
$4000 \lesssim \textrm{HJD} - 2450000 \lesssim 5200$, followed by a second, 
smaller-amplitude oscillation in the time interval $5500 \lesssim \textrm{HJD} - 2450000 \lesssim 5900$. 

\section{MONTE CARLO MICROLENSING AND TIME DELAY ANALYSIS}\label{sec:analysis_method}

As discussed in Section~\ref{sec:intro}, we wish to use our combined $r$- 
and $g$-band light curves to derive a more precise time delay for SBS\,0909
as well as to place new constraints on the quasar continuum source size and structure.
We do so here utilizing the alternate implementation of the \citet{kochanek04} microlensing
analysis techniques presented in \citet{morgan08,morgan12} in which the time delay and 
source size of a lensed quasar are simultaneously determined through a Bayesian
analysis of Monte Carlo microlensing simulations for a range of trial time delays. 
Our approach here differs only in that we analyze two optical light curves ($r$- and $g$-band) for each quasar
image rather than an optical light curve and an X-ray light curve for each image.
We carry out the microlensing and time delay analysis in two major steps: first, we 
analyze Monte Carlo simulations of the longer, better-sampled light curves (in this case, $r$-band) 
for a range of time delays and source sizes.  Then, we simultaneously analyze Monte Carlo simulations
of the $g$ and $r$ light curves incorporating the best time delay value and the continuum source 
size distribution produced in the first stage.  The final results for the $r$- and $g$-band 
continuum source sizes are derived in this second step.

\subsection{Time Delay Analysis of $r$-band Light Curves}\label{sec:delay}

We analyze only the $r$-band light curves for the time delay analysis, as the 
$r$-band monitoring spans a longer length of time and includes the largest-amplitude
flux variation.  To begin the analysis, we generate $r$-band light curve pairs 
in which image A's (the less variable image) light curve is shifted by a set 
of trial time delays, $\Delta t_{AB} = t_{A} - t_{B}$, according to the procedure
outlined in \citet{morgan08}.  Our set of trial time delays spans
the range $-70\,\textrm{days} \leq \Delta t_{AB} \leq 70\,\textrm{days}$ in time steps of
one day.  The shifting of the light curves
by the trial time delays must take into account two issues.  First, as explained in 
\citet{morgan08,morgan12}, each A/B light curve pair must have measurements for the
same dates, necessitating interpolation of the shifted image A measurements when they fall
in the middle of an observing season and extrapolation of image A measurements when the
shifted data points fall in interseason gaps.  We used linear interpolation for epochs falling
in the middle of an observing season and permitted 10 days of linear extrapolation
for points in interseason gaps, increasing the uncertainties for successive extrapolated 
points as described in \citet{morgan12}.  The second issue is 
that all trial light curves must have the same number of epochs, 
requiring the truncation of light curves for trial time delays shorter than 60 days.  
After all interpolations, extrapolations, and truncations, the shifted light curves used for 
our analysis each contained 158 epochs.
\begin{figure}
\begin{center}
\epsscale{1.2}
\plotone{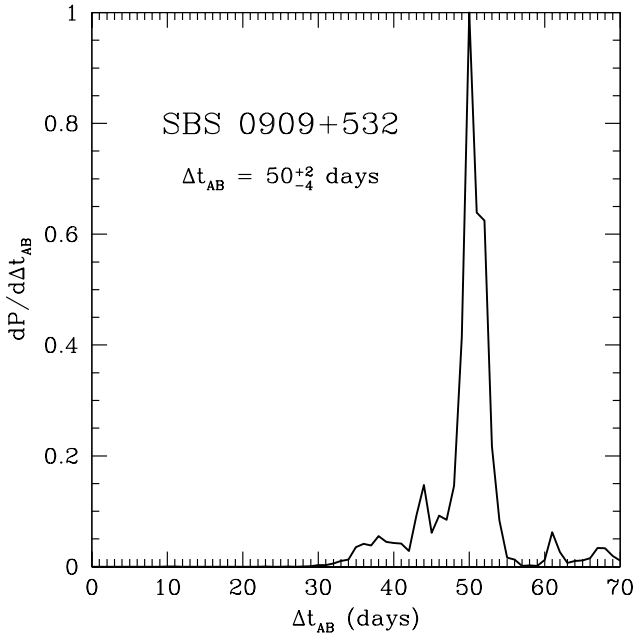}
\caption{Posterior probability distribution for the time delay in SBS\,0909.  The portion of the
distribution for delays $-70\,\textrm{days} < \Delta t_{AB} < 0\,\textrm{days}$ is not shown because the 
probability in that section is essentially zero.  Our result for the time delay, 
$\Delta t_{AB} = 50^{+2}_{-4}\,\textrm{days}$, where B leads A, agrees with the previous result from \citet{luis08},
but is more precise.\label{fig:time_delay_dist}}
\end{center}
\end{figure}

Next, for each trial time delay, we use the Monte Carlo method of 
\citet[][also see Poindexter \& Kochanek 2010]{kochanek04} to generate large numbers 
of light curves from microlensing magnification patterns for random 
combinations of effective velocity $v_{e}$ between quasar, lens galaxy, and observer,
mean microlens mass, and lens galaxy stellar mass/dark matter fraction.  We then 
fit the simulated light curves to the observed $r$-band light curve shifted by the trial time delay.  
To construct the magnification patterns, we must utilize physical models
of the macroscopic (strong) lensing, microlens mass function, and accretion disk
surface brightness profile.  We used the \emph{lensmodel} software package \citep{keeton01} to 
generate a sequence of strong lensing models for the SBS\,0909 system over a range of 
mass contributions from the dark matter and stellar components of the lens galaxy.  Each model is
a sum of a de Vaucouleurs component, representing the stellar content
of the lens galaxy, and a \citet[NFW;][]{navarro96} component, representing the galaxy's
dark matter halo and concentric with the de Vaucouleurs component.  We ran our first realization of the Monte Carlo 
simulation using a model sequence whose coefficients best reproduce the lens galaxy model of
\citet{lehar00}, but we eventually used a range of model sequences at both the Leh\'{a}r and \citet{sluse12} positions. 
A model sequence contains ten models spanning $0.1 \leq f_{M/L} \leq 1.0$
in steps of 0.1, where $f_{M/L}$ represents the ratio of the mass of the stellar component
to its mass in a uniform mass-to-light ratio model.  For the stellar (microlens) mass 
function, we use a power law, $dN/dM \propto M^{-1.3}$, with a ratio of
maximum-to-minimum mass ratio of 50; this function reasonably
approximates the Galactic disk mass function of \citet{gould00}. 
We model the quasar's accretion disk as a face-on, thin disk
radiating as a blackbody with a power-law temperature profile $T \propto R^{-3/4}$.
Our model matches the outer regions of the thin disk model of \citet{shakura73},
but we neglect the drop in temperature in the center due to the inner edge of the
disk and the correction factor from general relativity to avoid introducing additional
parameters.  Provided the disk sizes we obtain are significantly larger than the
radius of the inner disk edge, these simplifications introduce insignificant uncertainties 
relative to those associated with other parameters \citep{dai10}.
With these parameters we create 40 independent magnification patterns for each quasar image
for each of the ten different strong lens models, using the method described in the
Appendix of \citet{kochanek04}.  The patterns are 
$8192 \times 8192$ images representing $20\langle R_{E} \rangle \times
20 \langle R_{E} \rangle$, where $\langle R_{E}\rangle$ is the
Einstein radius for the mean microlens mass $\langle M \rangle$ projected
into the source plane, yielding a pixel scale of $1.1\times 10^{14}(\langle M \rangle/\msun)^{1/2}\,\textrm{cm}$.  
The outer dimensions and pixel scale are chosen to
be sufficiently large to representatively sample the magnification distribution and
sufficiently small to adequately resolve the accretion disk in the $g$-band simulations
(see Section~\ref{sec:dual_band}). 
\begin{figure*}
\epsscale{1.15}
\plottwo{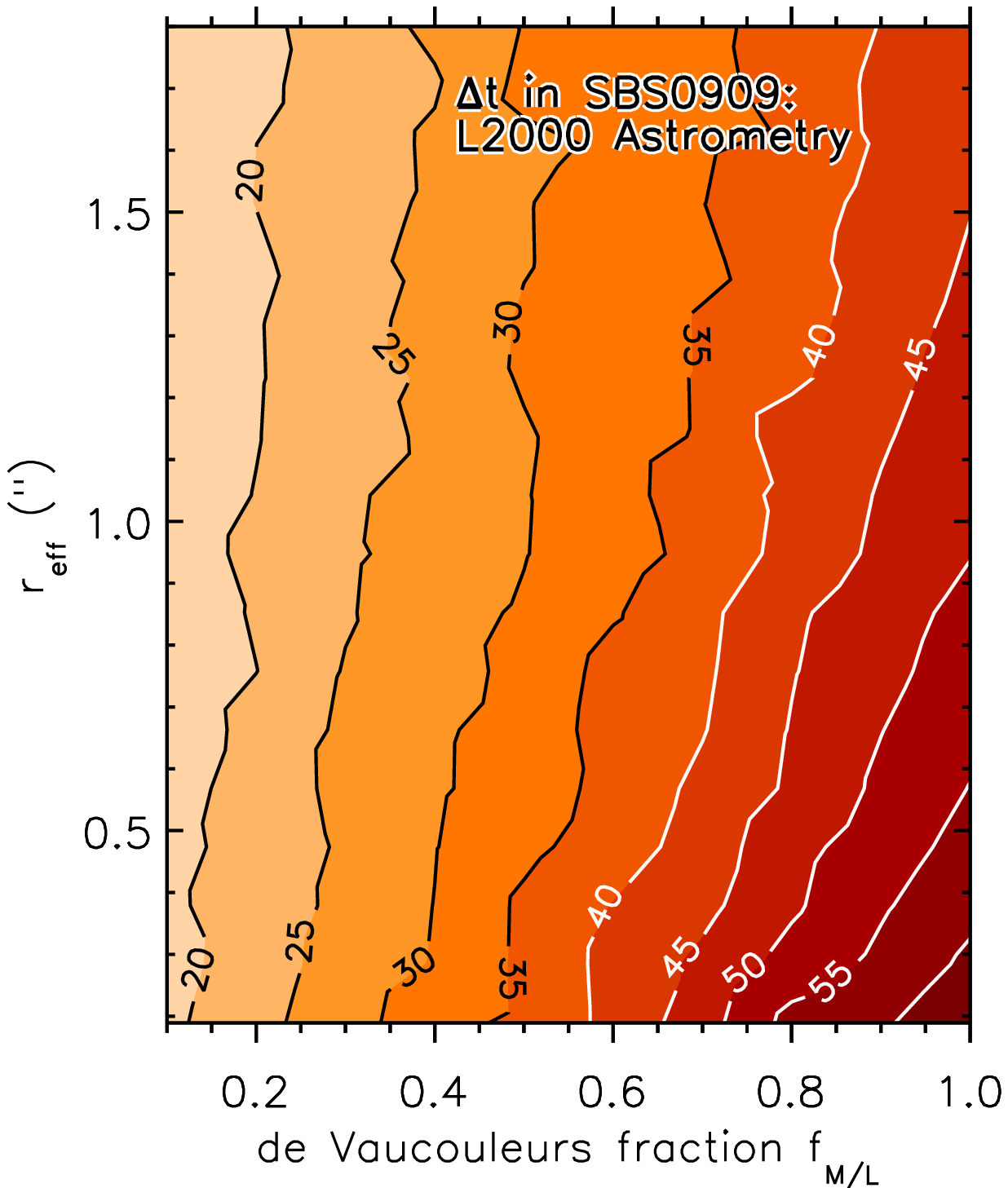}{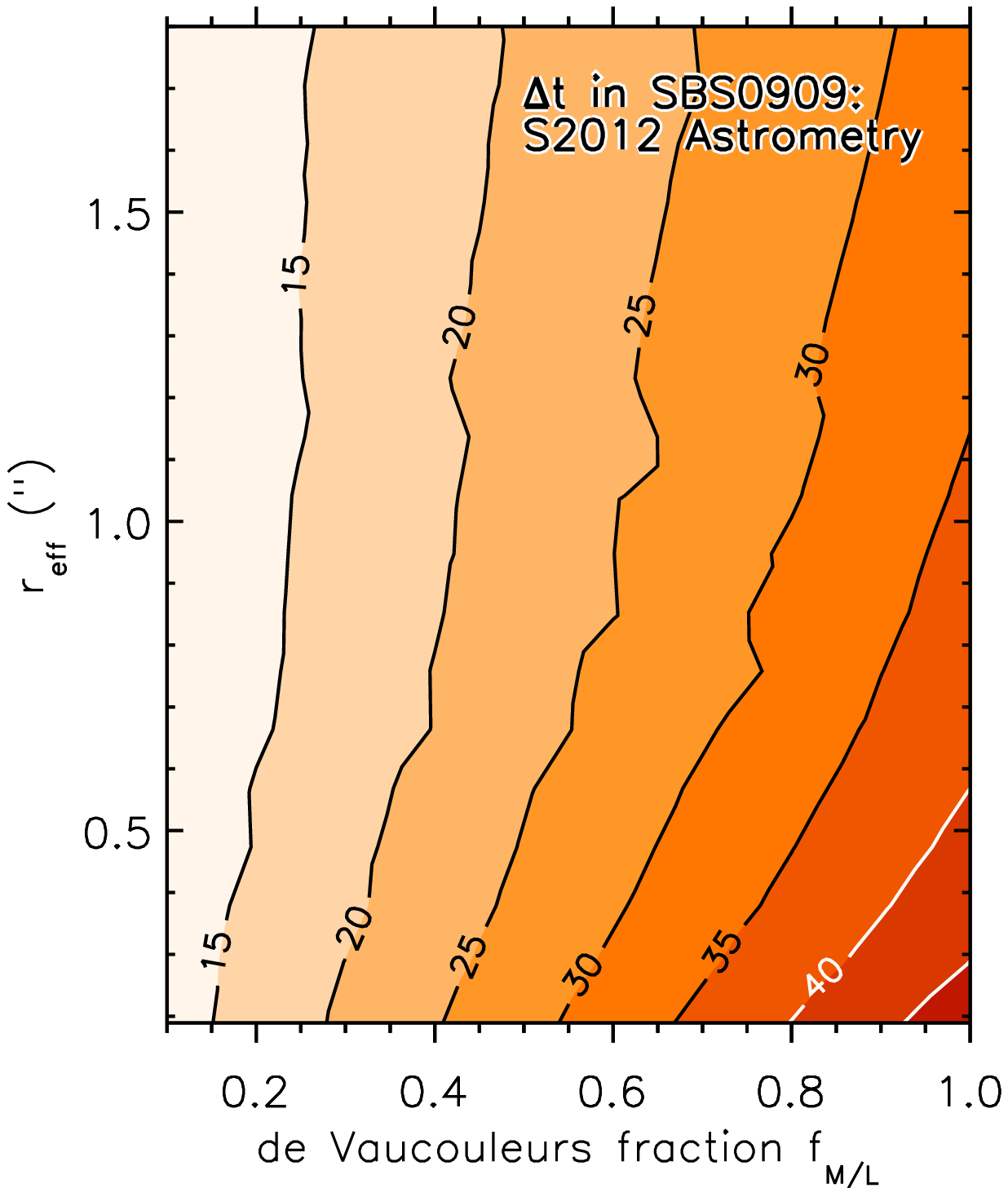}
\caption{Contours of time delay $\Delta t_{AB}$ predicted by a grid of 
de Vaucouleurs + NFW  lens models as a function of the deVaucouleurs fraction 
$f_{M/L}$ and effective radius $r_{\textrm{eff}}$ for the astrometric solutions of \citet{lehar00} (left panel) and 
\citet{sluse12} (right panel). The contour spacing is 5\,days.  
For each effective radius, the base model is a de Vaucouleurs-only 
model ($f_{M/L} = 1.0$) in which the the mass parameter is allowed to vary to yield the best fit. 
At each $r_{\textrm{eff}}$, the $f_{M/L}$ parameter (see Section~\ref{sec:delay}) varies in 
uniform steps between 1.0 and 0.1 relative to the base model.  
\label{fig:fml_plot}}
\end{figure*}

For the time delay analysis we carry out $10^{7}$ realizations of the $r$-band
light curve from each of the 400 sets of magnification patterns and for each of the 141 trial time delays.
We randomly select an initial position and effective velocity for the source
trajectory from their prior distributions under the assumption that these 
variables are independent and uniformly distributed.
For computational simplicity, we neglect the motion of the stars within the lens
galaxy and describe the observer's motion as the projection of the CMB dipole velocity
onto the lens plane, as done by \citet{kochanek04}.  We compare the simulated
light curves to the observed light curves and calculate the goodness-of-fit ($\chi^{2}$) statistics
for each, discarding trials with a $\chi^{2}$ statistic per degree of freedom ($\chi^{2}/\nu$) 
greater than 3.  

As we discussed in Section~\ref{sec:intro}, the time delay measurement in SBS\,0909 
is made more challenging by the uncertainties in the lens galaxy photometric model. Since \citet{lehar00} and
\citet{sluse12} disagree about the astrometric position and effective radius ($r_{\textrm{eff}}$) of the lens galaxy, we took steps to ensure 
the robustness of our time delay measurement in the resulting degenerate model parameter space.
First, we created a series of additional model sequences for a range of lens galaxy effective radii ($r_{\textrm{eff}}$) at the astrometric positions as determined by 
both \citet{sluse12} and \citet{lehar00}.
Each of these model sequences employs a two component (deVaucouleurs -- NFW) lens galaxy, where we vary the stellar mass fraction as described above, 
imposing ellipticity constraints on the lens galaxy from Ê\citet{sluse12} in the former case, and a small ellipticity of $1 - b/a = 0 \pm 0.08$ in the latter case. 
Then, for each new model sequence, we generated a new set of magnification patterns with which we repeated the full-scale Monte Carlo time delay analysis described above.  
For completeness, we also repeated this procedure for a Singular Isothermal  Ellipsoid (SIE) model at both the \citet{sluse12} and the \citet{lehar00} positions.  

Following completion of the Monte Carlo light curve simulations for all model sequences, we aggregated the results from all trials and performed a Bayesian
analysis of the $\chi^{2}$ statistics of the light curve fits.  We formally marginalize over the microlensing
variables of source size, microlens mass, and stellar mass fraction used to construct the 
magnification patterns, as well as the 
effective velocities from the Monte Carlo simulations, to 
calculate the posterior probability density for the time delay of SBS\,0909.  By aggregating the 
results from the model sequences at both the \citet{lehar00} and \citet{sluse12} positions, we effectively marginalize over
uncertainties in the lens galaxy photometric fits as well.
The trials with high $\chi^{2}$ which were thrown out would not contribute significantly
to the Bayesian integrals, so removing them does not affect our results.

We present the posterior probability density for the time delay 
$\Delta t_{AB}$ of SBS\,0909 resulting from our Bayesian analysis of the full set of model sequences
in Figure~\ref{fig:time_delay_dist}.  For ease of viewing we show 
only the portion of the distribution for $0\,\textrm{days} < \Delta t_{AB} < 70\,\textrm{days}$;
the values of the probability density for negative time delays were essentially zero.
The time delay distribution is narrowly peaked, with a median
$\Delta t_{AB} = 50$\,days (so image B leads image A) and a 68\% confidence interval of 
$46\,\textrm{days} < \Delta t_{AB} < 52\,\textrm{days}$.  Our result is consistent with,
but more precise than, the previous time delay measurement 
$\Delta t_{AB} = 49 \pm 6\,\textrm{days}$ by \citet{luis08}, 
based on the first two seasons of $r$-band monitoring
data used in this study.  We use this revised time delay for the remainder of 
our analysis, while acknowledging that we were unable to estimate the influence of all possible systematic errors 
in our measurement technique.  So, as is the case with any sophisticated measurement, it is possible that we underestimated the influence of systematic errors, and 
these unknown systematics may contribute to the discrepancy we describe in the next paragraph and explore in Section~\ref{sec:results_disc}.

In Figure~\ref{fig:fml_plot}, we illustrate the influence of the discrepancies in the lens galaxy photometric fits from the literature. We 
display contour plots of the time delay predicted by the SBS\,0909 lens models as a function of the 
effective radius $r_{\textrm{eff}}$ and fractional mass of the de Vaucouleurs component relative to constant $M/L$ model, $f_{M/L}$, 
for lens galaxy astrometry from both \citet{sluse12} and \citet{lehar00}. 
As expected, the predicted time delay is longer for the more
compact lens models with larger stellar mass components, but we also see that the delays are systematically longer for a galaxy located
at the \citet{lehar00} position than at the \citet{sluse12} position.  
This difference arises because the lens galaxy in the \citet{lehar00} fit
is closer to image A than in the \citet{sluse12} fit, yielding a larger gravitational delay in image A.  
In any case, neither the fiducial \citet{lehar00} model, nor the fiducial \citet{sluse12} model can reproduce 
our new measurement of the time delay $\Delta t_{AB}=50^{+2}_{-4}$\,days, but galaxy models at the Leh\'{a}r 
position and on the small end of the band of uncertainty in the Leh\'{a}r effective radius measurement ($r_ {\textrm{eff}} = 1\farcs58\pm0\farcs9$)
yield delays that are easily consistent with our new time delay measurement. On the other hand, 
reproducing our new delay measurement with models at the \citet{sluse12} position 
requires a lens galaxy that is signiÞcantly smaller than the already much more compact 
galaxy in Sluse measurement ($r_{\textrm{eff}}=0\farcs54 \pm 0\farcs02$).
Fortunately, our Monte Carlo microlensing simulation is sufficiently realistic as to be sensitive to the 
differences between intrinsic and microlensing variability, despite the uncertainties in the macroscopic lens model, since for each model sequence 
we sample a wide range of the stellar-to-total convergence ratio $\kappa_*/\kappa$, with significant overlap between the macro model sequences.

\subsection{Simultaneous Dual-Band Microlensing Analysis}\label{sec:dual_band}

Our dual-band ($r$ and $g$) microlensing analysis follows the method of the
simultaneous optical and X-ray analyses by \citet{dai10} and \citet{morgan12}.  We first shift
the $r$ and $g$-band light curves by the new time delay, $\Delta t_{AB} = 50\,\textrm{days}$,
in the same manner as we employed to construct the difference light curves in 
Section~\ref{sec:construct_lightcurves}.
Using the same magnification patterns from the time delay analysis, we carry out $10^{7}$
simulations of the $r$-band light curve for each of the sets of magnification patterns and
discard solutions for which $\chi^{2}/\nu > 2.5$.  
We saved all the physical parameters from the surviving $r$-band light curve fits. 
We then attempted to fit the $g$-band light curve using the trajectories from the best $r$-band fits for a new
grid of source sizes, and we compute the joint $\chi^{2}/\nu$ for the combined $r$- and $g$-band fits. 
A sample simulated $r$- and $g$-band
difference light curve which is a best fit to the observed data is shown in 
Figure~\ref{fig:diff_lightcurve_fits}.

We calculate posterior probability density distributions for the accretion disk sizes 
in the $g$- and $r$-bands and the lens galaxy stellar mass fraction ($f_{M/L}$)
by performing a Bayesian analysis on the combined set of $r$- and
$g$-band solutions.  We note that our simulations are carried out in Einstein units, where
source sizes and velocity are scaled by $(\langle M/\msun \rangle)^{1/2}$
and denoted by $\hat{r}_{s}$ and $\hat{v}_{e}$, respectively.  To obtain
the probability density for the true, unscaled physical source size $P(r_{s})$
from that for the scaled source size, $P(\hat{r}_{s})$, we combine $P(\hat{r}_{s})$ with the 
probability density for the scaled effective velocity, $P(\hat{v}_{e})$, and
a statistical model (i.e., a prior) for the true effective source velocity, $P(v_{e})$, in
our analysis.  We construct $P(v_{e})$ using the method described in \citet{kochanek04}. 
For that purpose we use the peculiar velocity estimates for the redshifts of
SBS\,0909 and the lens galaxy from the models presented in \citet{mosquerakochanek11}
and estimate the velocity dispersion of the lens galaxy from its Einstein radius, 
assuming the galaxy is a singular isothermal sphere with relaxed dynamics, which
\citet{treu04} and \citet{bolton08} show is a good approximation. 
As a final step we must correct the scale radius for the disk's inclination $i$ by 
multiplying by $(\cos i)^{-1/2}$, which is necessary because we have assumed a 
face-on disk in our simulations and microlensing amplitudes depend on the 
projected area of a source rather than the shape.  

\section{RESULTS \& DISCUSSION}\label{sec:results_disc}

\begin{figure}
\epsscale{1.2}
\plotone{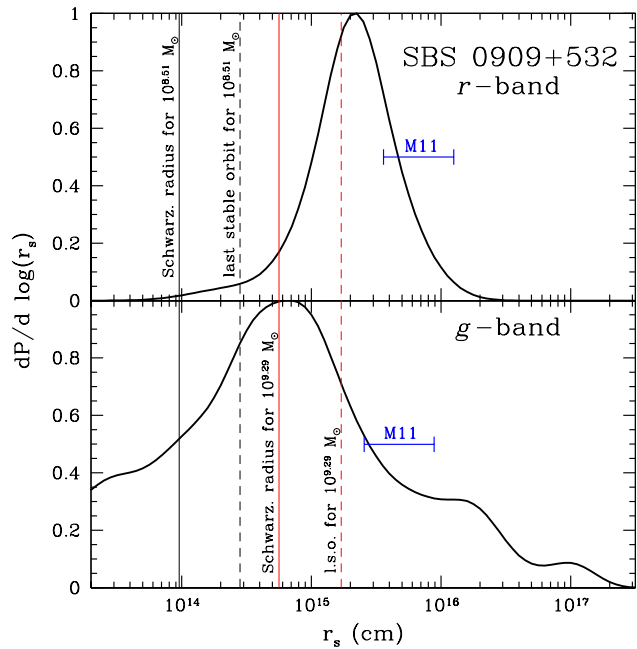}
\caption{Joint probability distributions for physical scale size of the accretion
disk in observed-frame $r$ (top panel) and $g$ (bottom panel) bands in SBS\,0909.  Both
distributions have been corrected for inclination assuming $i=60\degr$.  The solid and 
dashed vertical lines indicate the Schwarzschild radius and the radius of the last stable orbit in the Schwarzschild 
metric, respectively, for black holes of mass $10^{8.51}\,\msun$ and $10^{9.29}\,\msun$.  Our
disk sizes are more consistent with a central black hole mass of $10^{8.51}\,\msun$ for SBS\,0909.
For comparison we show the $1\sigma$ range for the disk size of SBS\,0909 obtained
by \citet{mediavilla11}, scaled to our rest-frame wavelengths and for mean microlens mass
$\langle M \rangle = 0.3\,\msun$.  Our disk scale radii are marginally consistent with 
M11's result. \label{fig:source_size_plot}}
\end{figure}

In the top panel of Figure~\ref{fig:source_size_plot} we show the posterior probability 
distribution for the physical size of the quasar's accretion disk in the observed-frame $r$-band
resulting from our two-band Bayesian microlensing analysis.  Hereafter, we 
state all sizes in terms of the thin disk scale radius, $r_{s}$, defined as
the radius at which the disk temperature matches the rest-frame wavelength
of the filter used in our monitoring observations, $kT = hc/\lambda_{\textrm{rest}}$
(for $r$-band monitoring of SBS\,0909, $\lambda_{\textrm{rest}} = 2620\,\textrm{\AA}$; for
$g$-band, $\lambda_{\textrm{rest}} = 2020\,\textrm{\AA}$).  This
can be converted to a half-light radius using the relation $r_{1/2} = 2.44\,r_{s}$.
We note that the distribution shown and
all numerical quantities in the discussion which follows have been corrected for disk
inclination $i$ through multiplication by a factor of $\langle \cos{i} \rangle^{-1/2}$, assuming $i=60\degr$
(corresponding to the expectation value of a random distribution of disk inclinations).
The median of the probability distribution for $r$-band source size is 
$\log (r_{s,r}/\textrm{cm}) = 15.3 \pm 0.3$, where the error bar represents the bounds of the 
68\% confidence interval.  In Figure~\ref{fig:source_size_plot} we also show for comparison
the $1\sigma$ range of values for the accretion disk size obtained by \citet[][hereafter M11]{mediavilla11}
from their chromatic microlensing analysis of SBS\,0909 under the assumption of a logarithmic size prior
and a disk temperature profile power-law index $p = 1/\alpha = 4/3$.  In order to make the most accurate
comparison, we converted M11's half-light radius to a thin-disk scale size, scaled the result 
from its rest-frame wavelength of $1460\,\textrm{\AA}$ to $\lambda_{rest}(r) = 2620\,\textrm{\AA}$
assuming $R_{\lambda} \propto \lambda^{4/3}$ for thin disks, scaled once more to
a mean microlens mass $\langle M \rangle = 0.3\,\msun$, and corrected 
for inclination assuming the same $i=60\degr$ we applied to our disk sizes.  
As can be seen in Figure~\ref{fig:source_size_plot}, our microlensing source size 
for SBS\,0909 is smaller, but marginally consistent with M11's. We suspect that the
difference may arise from the effect of the magnification pattern pixel sizes
on the size distribution (noting that our $r$-band disk size is similar to the size of the 
pixels in M11's magnification patterns), evidence of microlensing of the quasar's broad 
emission lines, and the different modeling approaches used in the different studies.  
It is also conceivable that M11's result has been affected by their use of single-epoch 
spectra and/or their combination of optical and near-IR spectra obtained at epochs separated
by several years: the uncorrected time delay and intrinsic variability may alter 
the continuum and emission line flux ratios from their true values.

Because it is possible that the $r$-band flux which we observe from SBS\,0909 and model in our simulations
could be contaminated by UV or optical photons scattered by the broad line region, or higher energy continuum
photons reprocessed by the broad line region and re-emitted as emission lines, our $r$-band accretion
disk size may be an overestimate \citep[see][]{morgan10,guerras13}.  In fact, the prominent \ion{Mg}{2} 
emission line (rest-frame 2798\,\AA) in the spectrum of SBS\,0909 
falls within the passband of our $r$-band filter \citep[see][M11]{lubin00}, 
so contamination from line emission is of particular concern.  To investigate the possibility, 
we have repeated our microlensing simulations under the assumption that a fraction of 
the observed $r$-band flux should actually be attributed to unmicrolensed
emission from scattered light or the broad line region.  We find, however, that unmicrolensed contamination is
not a significant factor in our accretion disk size determination: even when we assume that as much as
30\% of the observed $r$-band flux is contributed by contamination from
emission on large physical scales, the median of the $r$-band size
probability distribution is essentially unchanged at $\log (r_{s,r}/\textrm{cm}) = 15.3^{+0.3}_{-0.4}$.
\begin{figure}
\epsscale{1.2}
\plotone{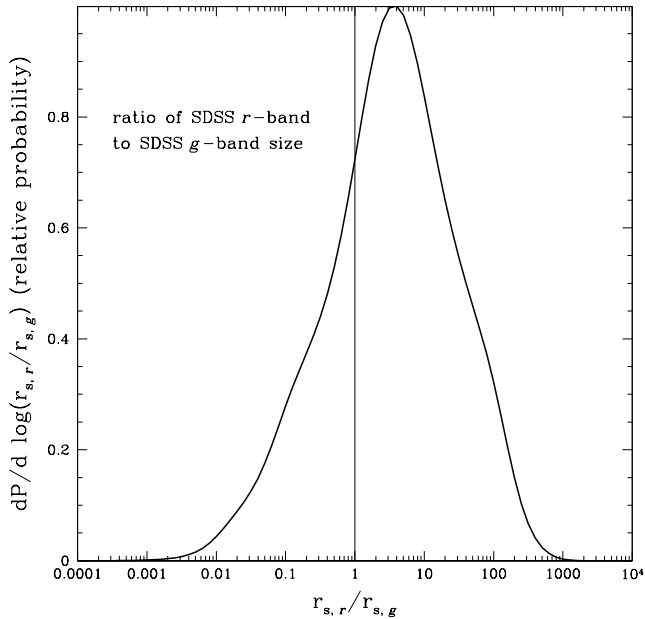}
\caption{Joint probability density for the ratio of the accretion disk sizes in observed-frame
$r$-band and $g$-band ($r_{s,r}/r_{s,g}$) for SBS\,0909.  The vertical line highlights the 
location of $r_{s,r}/r_{s,g} = 1$.  The distribution is very wide, reflecting the poor constraints
we are able to place on the observed-frame $g$-band accretion disk size.  The median and 1$\sigma$
values for the size ratio distribution are $\log{r_{s,r}/r_{s,g}} = 0.5^{+0.9}_{-1.0}$, which are larger but not statistically
inconsistent with the $r/g$-band size ratio expected for a thin accretion disk. \label{fig:size_ratio}}
\end{figure}

We show the posterior probability density for the observed-frame $g$-band accretion disk
size for SBS\,0909 resulting from our two-band microlensing analysis in the lower panel of 
Figure~\ref{fig:source_size_plot}.  We have included in this panel the $1\sigma$ range 
of the accretion disk size result from M11 as well, corrected as described in the paragraph above, 
except the result has now been scaled to the rest-frame wavelength of $g$-band ($2020\,\textrm{\AA}$)
instead of $r$-band.  As might be expected due to the shorter time baseline of the $g$-band 
monitoring data and somewhat poorer data quality, the constraints we obtain for the $g$-band
size are not nearly as tight as those for $r$-band.  In light of the wide peak of
the $g$-band probability distribution and the hints of secondary peaks, we regard the $g$-band result
as a preliminary estimate.  Despite the large uncertainty, it is encouraging that the median 
of the $g$-band physical size distribution, $\log (r_{s,g}/\textrm{cm}) = 14.8 \pm 0.9$, is indeed smaller than 
the median of our $r$-band size distribution, consistent with the shorter wavelength of $g$-band
and the values for the disk temperature slope in the literature.  Like our $r$-band disk size,
our $g$-band result is notably smaller than the scaled result from M11, although the 
significance of the discrepancy is low due to the large uncertainties.
We also calculate the probability density for the ratio of the $r$- and $g$-band disk sizes, 
which we show in Figure~\ref{fig:size_ratio}.
The distribution is very broad, with a median value and $1\sigma$ confidence level of 
$\log{r_{s,r}/r_{s,g}} = 0.5^{+0.9}_{-1.0}$.  Using a standard thin-disk temperature profile
($T\propto R^{-3/4}$) to predict the observed-frame $r$/$g$ size ratio would result in 
$\log{r_{s,r}/r_{s,g}} = 0.15$, which is smaller but statistically consistent with
our observed value.  Unfortunately, our best value for the $r$/$g$ size ratio is too uncertain to 
provide any conclusive indication of the temperature profile of the accretion disk in SBS\,0909. 
We expect that a future analysis utilizing a longer $g$-band time baseline for photometric monitoring
will significantly improve the precision of the $g$-band size measurement.

An interesting point made by Figure~\ref{fig:source_size_plot} is that the accretion disk sizes we
obtain from our two-band microlensing analysis are not consistent with the central supermassive black hole mass
derived from SBS\,0909's H$\beta$ emission lines by \citet{assef11}, $10^{9.29}\,\msun$.
The H$\beta$ line measurement appears to be the most reliable for SBS\,0909 in \citet{assef11}, 
because the line profiles for \ion{C}{4} and H$\alpha$ emission lines were difficult to model. 
By extension, we might expect the black hole mass calculation from H$\beta$ to be the most reliable as well.  
However, the innermost stable orbit of a maximally rotating Kerr black hole and the Schwarzschild radius predicted
for a black hole of mass $10^{9.29}\,\msun$ both fall within
the 1$\sigma$ bounds of our $r$- and $g$-band accretion disk sizes.  
Another possible clue that the H$\beta$ black hole mass may 
be problematic is that the theoretical scale radius of a thin accretion disk at 2620\,\AA\ 
surrounding a black hole of mass $10^{9.29}\,\msun$ (the ``theory size", 
$\log [R_{2620}/\textrm{cm}] = 15.73$) is larger
than the $r$-band microlensing size from our simulations; yet, in the quasar microlensing literature,
the theory size is consistently smaller than the results of microlensing simulations 
\citep[see, e.g.,][]{morgan10,blackburne11}.  The accretion disk sizes 
predicted by our microlensing simulations are more consistent with the \ion{C}{4} black hole mass estimate, 
$10^{8.51}\,\msun$, highlighting the difficulties and uncertainties associated with estimating
black hole masses from quasar emission lines.

Our Monte Carlo microlensing analysis of SBS\,0909 $g$- and $r$-band monitoring data has 
enabled us to estimate the $r$-to-$g$-band accretion disk size ratio and improve the precision of the system's time delay,
despite uncertainties in the macroscopic lens galaxy model.  We suggest that deep, high-resolution 
imaging of the SBS\,0909 system will be necessary to completely resolve the lingering questions about the lens galaxy model. 
For the moment, our delay measurement leads us to favor the lens galaxy astrometry of \citet{lehar00} over that of \citet{sluse12}.
Our result for the $r$/$g$-band size ratio is rather coarse and requires 
confirmation; to that end, we have begun monitoring
SBS\,0909 in $g$-band again.  Additionally, we recently expanded the USNA/USNO lensed quasar monitoring campaign 
to near-infrared (NIR) wavelengths.  We look forward to the improvements in our ability to constrain the temperature 
profiles of quasar accretion disks that will be enabled by size measurements across a significantly larger
wavelength baseline. 

\acknowledgements

This material is based upon work supported by the National Science Foundation under
grants No.\ AST-0907848 and AST-1211146 (to C.W.M.), and AST-1009756 (to C.S.K.).
C.W.M. also gratefully acknowledges support from the Research
Corporation for Science Advancement and Chandrasekhar X-Ray Center award 11700501.
The Liverpool Telescope is operated on the island of La Palma by Liverpool John Moores 
University in the Spanish Observatorio del Roque de los Muchachos of the 
Instituto de Astrof\'{\i}sica de Canarias with financial support from the UK Science and 
Technology Facilities Council. The Liverpool Quasar Lens Monitoring (LQLM) program is 
supported by the Spanish Department of Science and Innovation grant AYA2010-21741-C03-03 
(Gravitational LENses and DArk MAtter - GLENDAMA project), and the University of Cantabria.

\end{document}